\documentclass[aps,prl,twocolumn,showpacs,superscriptaddress]{revtex4}
\usepackage{graphicx}
\usepackage{dcolumn}
\usepackage{bm}

\newcommand{\be}{\begin{equation}}
\newcommand{\ee}{\end{equation}}
\newcommand{\bea}{\begin{eqnarray}}
\newcommand{\eea}{\end{eqnarray}}
\newcommand{\beas}{\begin{eqnarray*}}
\newcommand{\eeas}{\end{eqnarray*}}
\newcommand{\bes}{\begin{equation*}}
\newcommand{\ees}{\end{equation*}}

\begin{document}

\title{Demonstration of a Quantum Controlled-{\sc not} Gate in the Telecom Band}

\author{Jun Chen}
\affiliation{Center for Photonic Communication and Computing, EECS Department\\
Northwestern University, 2145 Sheridan Road, Evanston, IL
60208-3118}
\author{Joseph B. Altepeter}
\affiliation{Center for Photonic Communication and Computing, EECS Department\\
Northwestern University, 2145 Sheridan Road, Evanston, IL
60208-3118}
\author{Milja Medic}
\affiliation{Center for Photonic Communication and Computing, EECS Department\\
Northwestern University, 2145 Sheridan Road, Evanston, IL
60208-3118}
\affiliation{Department of Physics and Astronomy,
Northwestern University, 2145 Sheridan Road, Evanston, IL
60208-3118}
\author{Kim Fook Lee}
\affiliation{Center for Photonic Communication and Computing, EECS Department\\
Northwestern University, 2145 Sheridan Road, Evanston, IL
60208-3118}
\author{Burc Gokden}
\affiliation{Center for Photonic Communication and Computing, EECS Department\\
Northwestern University, 2145 Sheridan Road, Evanston, IL
60208-3118}
\author{Robert H. Hadfield}
\affiliation{National Institute of Standard and Technology, 325
Broadway, Boulder, CO 80305}
\author{Sae Woo Nam}
\affiliation{National Institute of Standard and Technology, 325
Broadway, Boulder, CO 80305}
\author{Prem Kumar}
\affiliation{Center for Photonic Communication and Computing, EECS Department\\
Northwestern University, 2145 Sheridan Road, Evanston, IL
60208-3118} \affiliation{Department of Physics and Astronomy,
Northwestern University, 2145 Sheridan Road, Evanston, IL
60208-3118}

\begin{abstract}
We present the first quantum controlled-\textsc{not}
(\textsc{cnot}) gate realized using a fiber-based
indistinguishable photon-pair source in the 1.55\,$\mu$m
telecommunications band. Using this free-space \textsc{cnot} gate,
all four Bell states are produced and fully characterized by
performing quantum state tomography, demonstrating the gate's
unambiguous entangling capability and high fidelity. Telecom-band
operation makes this \textsc{cnot} gate particularly suitable for
quantum information processing tasks that are at the interface of
quantum communication and linear optical quantum computing.
\end{abstract}


\maketitle

Photons in the telecommunications band (e.g., 1.55\,$\mu$m
wavelength) are ideally suited for carrying quantum information
over a large-scale quantum network, because such a network could
use low-loss optical fibers. Moreover, photons in optical fibers
interact weakly with their surrounding environments, displaying
low decoherence. These desirable features, together with easily
realizable single qubit operations, make the telecom-band photons
a prominent candidate for various quantum-information-processing
applications~\cite{Nielsen}. The main obstacle to using photons
for quantum computing tasks---including the most basic quantum
computing operation, the controlled-\textsc{not} (\textsc{cnot})
gate---is the miniscule interaction between two photonic `qubits'
(quantum bits). Fortunately, this obstacle has been circumvented
by the arrival of a seminal paper~\cite{KLM} in which the required
nonlinearity between photons has been effectively transferred to
measurement and post-selection.

This Letter describes efforts to apply this new quantum
computational paradigm to the fundamental \textsc{cnot} operation
using telecom-band photons. Previous implementations of the
\textsc{cnot} gate (or a similar controlled-phase
gate)~\cite{Pittman,Obrien,Langford,Kiesel,Sasaki,Zeilinger,Zhao}
utilized photons from spontaneous parametric down-conversion
(SPDC) in second-order ($\chi^{(2)}$) nonlinear crystals. Those
SPDC photons are not telecom band, and are therefore subject to
much higher losses when transferred through optical fiber.
Although telecom-band photons can also be generated via the SPDC
process, these photons are naturally emitted into a large number
of spatial and spectral modes, resulting in significant losses
when coupled to a single-mode optical fiber. The usage of SPDC
photon sources in fiber quantum networks is thus limited. However,
recently developed fiber-based sources~\cite{ChenPRL,ChenOL}
intrinsically avoid this issue. In this Letter we demonstrate, for
the first time to the best of our knowledge, a quantum
\textsc{cnot} gate at a telecom-band wavelength. This gate uses
three separate yet individually crucial experimental components: a
fiber-based indistinguishable photon-pair source~\cite{ChenPRL},
the free-space linear optical components of the \textsc{cnot} gate
itself, and a heralding detection system composed of one
superconducting single-photon detector (SSPD) and one InGaAs/InP
avalanche photodiode (APD).

\begin{figure}
\centering
\includegraphics[scale=0.35]{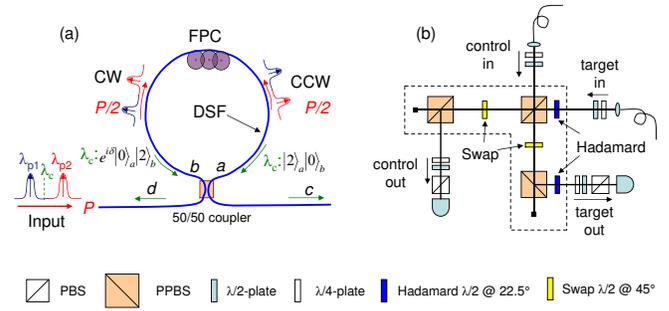} \vspace{-18pt}
\caption{(color online) Key components of the \textsc{cnot} gate.
(a) Quantum splitter (QS) source. Two identical photon-pair
wavefunctions interfere at a 50/50 beam splitter with zero phase
difference, resulting in deterministic splitting of the photon
pair. DSF, dispersion-shifted fiber; FPC, fiber polarization
controller; CW, clockwise; CCW, counterclockwise. (b) Structure of
the quantum \textsc{cnot} gate. (P)PBS, (partially) polarizing
beam splitter.} \label{QSCNOT} \vspace{-12pt}
\end{figure}

Indistinguishable (or {\it identical}) photons refer to photons
having the same spatial, temporal (frequency) and polarization
mode structure, and constitute a critical resource for linear
optical quantum computing. Quantum interference arising from the
bosonic nature of indistinguishable photons, such as the
well-known Hong-Ou-Mandel (HOM) interference~\cite{HOM}, lies at
the heart of linear optical quantum computing. Our group has
recently demonstrated a HOM dip of approximately 94\% visibility
(without any accidental subtraction) using a fiber-based
indistinguishable photon-pair source~\cite{ChenPRL}, which we
dubbed a ``quantum splitter", or QS for short. The QS source is
shown in Fig.~\ref{QSCNOT} (a), and can be conveniently summarized
in terms of ``time-reversed Hong-Ou-Mandel interference". To see
how, consider a dual-frequency pump~\cite{ChenOL}, which consists
of two copolarized, temporally overlapped, and spectrally distinct
(center wavelengths $\lambda_{p1}$ and $\lambda_{p2}$) pulses with
equal power, entering a fiber Sagnac loop from port $d$. The
Sagnac loop is composed of a 50/50 fiber coupler, a piece of
dispersion-shifted fiber (DSF), and a fiber polarization
controller (FPC). The total peak pump power $P$ is equally split
into two dual-frequency, counter-propagating pump pulses of power
$P/2$. The DSF inside the Sagnac loop is chosen such that its
zero-dispersion wavelength ($\lambda_0$) is close to the mean
wavelength ($\lambda_c$) of the pump's two central wavelengths
[i.e., $\lambda_0 \simeq \lambda_c \equiv
2\lambda_{p1}\lambda_{p2}/(\lambda_{p1}+\lambda_{p2})$] to
maximize the four-wave mixing (FWM) scattering
efficiency~\cite{Agrawal}. The FWM process of interest is of the
reverse degenerate type~\cite{Fan,ChenOL}, wherein two pump
photons of different frequencies ($\omega_{p1}$ and $\omega_{p2}$)
annihilate to produce a pair of energy-degenerate daughter photons
at their mean frequency ($\omega_c$), satisfying
$\omega_{p1}+\omega_{p2}=2\omega_c$. When their powers are
balanced, the clockwise pump and the counterclockwise pump scatter
copolarized FWM photon-pairs with equal probability. The two
probability amplitudes are then made to interfere at the 50/50
fiber coupler; their phase difference $\delta$ is controlled by
the setting of the intraloop FPC. For an input state
$|\Psi\rangle_{\rm in} = (|2\rangle_a |0\rangle_b + e^{i \delta}
|0\rangle_a |2\rangle_b)/\sqrt{2}$, the output state for a
standard symmetric beam splitter is given by $|\Psi\rangle_{\rm
out} = (1-e^{i \delta})\, \Psi_{\overline{2002}}/2 + i (1+e^{i
\delta})\, \Psi_{11}/2$, where
$\Psi_{\overline{2002}}\equiv(|2\rangle_d |0\rangle_c -
|0\rangle_d |2\rangle_c)/\sqrt{2}$ and $\Psi_{11}\equiv|1\rangle_c
|1\rangle_d$. It is transparent from the above expressions that if
we set $\delta$ to be 0, we will perform the reverse operation
($\Psi_{2002} \Rightarrow \Psi_{11}$) of the conventional HOM
interference ($\Psi_{11} \Rightarrow \Psi_{2002}$) with
$\Psi_{2002}\equiv(|2\rangle_a |0\rangle_b + |0\rangle_a
|2\rangle_b)/\sqrt{2}$, and hence the interpretation as
``time-reversed HOM interference". The name ``quantum splitter",
however, stems from the intended use of such a device to {\it
deterministically} split two identical photons. Note that in this
QS setting, the Sagnac loop also splits the classical pump into
two equal-powered components in ports $c$ and $d$.

The output of the QS source is fed into a \textsc{cnot} gate based
on three partially polarizing beam splitters (PPBS), as shown in
Fig.~\ref{QSCNOT} (b). Requiring only two input photons (control
and target) with no ancillary photons, this simple {\sc cnot} gate
is probabilistic in nature~\cite{Langford,Sasaki,Kiesel}; however
it is still in principle scalable when coupled with linear optical
quantum non-demolition measurements~\cite{Obrien}. The PPBSs
adopted in the {\sc cnot} gate design are optical devices which
completely reflect vertically-polarized light ($V$), and have a
reflectivity of 1/3 for horizontally-polarized light ($H$).
Non-classical interference of the HOM type happens only at the
central PPBS, and only for horizontally-polarized input qubits
($HH$, where the first $H$ refers to the control qubit and the
second $H$ refers to the target qubit). The other two PPBSs, each
preceded by a swap gate (a half-wave plate with its principle axis
set to $45^\circ$, performing $H\leftrightarrow V$), exist to
equalize the probability amplitudes for all other inputs ($VV$,
$VH$ and $HV$). With the logic-basis definitions $0 \equiv V$ and
$1 \equiv H$, it is a simple exercise to show that the gate
succeeds with a probability of $1/9$ in performing the following
\textsc{cnot} transformation: $\alpha |VV\rangle$ + $\beta
|VH\rangle$ + $\gamma |HV\rangle$ + $\delta |HH\rangle$
$\rightarrow$ $\alpha |VV\rangle$ + $\beta |VH\rangle$ + $\gamma
|HH\rangle$ + $\delta |HV\rangle$, conditioned on the detection of
one and only one photon in each of the output modes.


\begin{figure}
\centering
\includegraphics[scale=0.35]{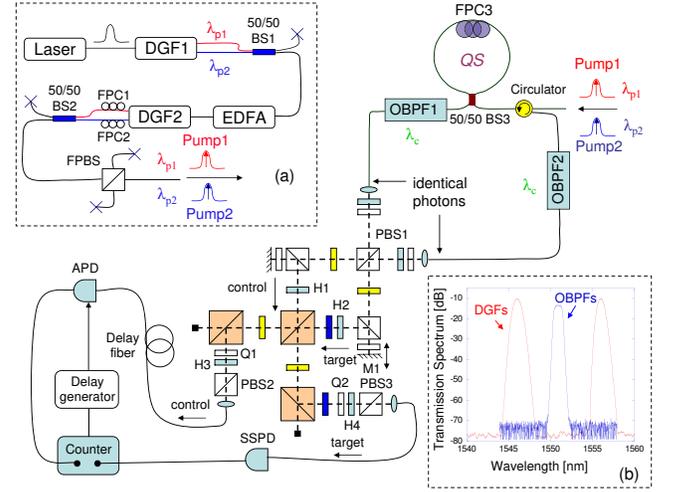} \vspace{-18pt}
\caption{(color online) Schematic experimental setup. Identical
photons generated by the QS source serve as input control and
target qubits to the \textsc{cnot} gate, whose output is collected
using a heralding detection scheme with a superconducting
single-photon detector (SSPD) and an avalanche photodiode (APD).
BS, beam splitter; OBPF, optical bandpass filter. Inset (a):
Preparation of the dual-frequency copolarized pump. DGF,
double-grating filter; FPBS, fiber polarizing beam splitter; EDFA,
erbium-doped fiber amplifier. Inset (b): Transmission spectra of
the OBPFs and DGFs.} \label{Expr} \vspace{-12pt}
\end{figure}

Figure~\ref{Expr} depicts the experimental setup for demonstrating
the \textsc{cnot} gate's operation. We pump the QS source with a
dual-frequency copolarized pump, which is obtained from spectral
carving of a broadband mode-locked femtosecond laser (repetition
rate $\simeq 50$\,MHz). The details of the dual-frequency pump
have been described in Ref.~\cite{ChenOL}, and is shown in inset
(a) for completeness. To recapitulate, we select our two pump
central wavelengths ($\lambda_{p1}=1545.95$\,nm and
$\lambda_{p2}=1555.92$\,nm, pulse width $\simeq 5$\,ps) by passing
the broad laser spectrum through two double-grating filters [DGF1
and DGF2, FWHM $\simeq 0.8$\,nm for each passband; see inset (b)
in Fig.~\ref{Expr} for their spectral shape]. An erbium-doped
fiber amplifier is sandwiched in between the two DGFs to provide
pump power variability. Each pump pulse's timing, polarization,
and power can be individually controlled, so that at the output of
the fiber polarizing beam splitter we obtain copolarized,
equal-powered, dual-frequency pump pulses with overlapped timing.
We then direct this pump toward the QS source. A circulator is
placed before the Sagnac loop, which redirects the
Sagnac-reflected photons to a separate spatial mode. A pair of
optical bandpass filters (OBPF1 and OBPF2) with identical
transmission spectrum [center wavelength $\lambda_c= 1550.92$\,nm,
passband FWHM $\simeq 0.8$\,nm, see inset (b) of Fig.~\ref{Expr}]
are utilized at the output ports of the Sagnac loop to select the
spatially separated identical photons at $\lambda_c$. The OBPFs
also provide the necessary $> 100$\,dB isolation from the pump in
order to effectively detect those filtered photons. Before pumping
with the dual-frequency pump, the 300\,m-long DSF in the Sagnac
loop is immersed in liquid nitrogen to suppress spontaneous Raman
scattering~\cite{Takesue,Lee}, and we ensure that thermal
equilibrium is reached between the DSF and its liquid-nitrogen
environment. We then align the Sagnac loop to its QS setting by
using a continuous-wave tunable laser set at wavelength
$\lambda_c$ as the input, and adjusting FPC3 so that the output
powers from OBPF1 and OBPF2 are proportional to each filter's
individual transmission efficiency~\cite{ChenPRL}.


Coming out of the QS source, the two identical photons are
collimated into free space, and directed through several optical
elements.
The goal of these optics is to precisely control the timing and
polarization of each photon, so that they are maximally overlapped
when they interfere nonclassically at the central PPBS of the
\textsc{cnot} gate. A translation stage is placed under the mirror
M1 to precisely match the paths of the two photons.
The first polarizing beam splitter (PBS1) then separates the two
photons into separate paths.
The other two PBSs, each with a $45^\circ$-quarter-wave plate and
a mirror behind, function just like two perfectly reflecting
mirrors. This
configuration ensures that translating M1
does not cause misalignment of the input target photon.
The control and target photons enter the \textsc{cnot} gate with
well-defined polarization (vertical) and timing (within their
$\simeq 5$\,ps pulse duration). Half-wave plates H1 and H2 are
used to define the logical values of input qubits, while two
polarization analyzers (Q1/H3/PBS2 and Q2/H4/PBS3) are employed at
the output to examine the polarization of the output qubits.

One of the primary obstacles to the implementation of telecom-band
quantum information protocols is the lack of good single-photon
detectors, in contrast to the high quality single-photon detectors
available for visible wavelengths.  For this experiment, we had
access to both cryogenically cooled SSPDs and InGaAs APDs. The
SSPDs~\cite{Nam2,Liang} have low efficiency ($\simeq 1\%$) and low
dark-count probability ($\simeq 3 \times 10^{-6}$ counts/gate).
The APDs (Epitaxx, EPM 239BA) have higher efficiency ($\simeq
20\%$) and much higher dark-count probability ($\simeq 3 \times
10^{-3}$ counts/gate). (These rates should be compared with the
inferred FWM production rate in this experiment---$\simeq 0.15$
FWM photon pairs/gate and $0.15^{2}$ multiple-photon pairs/gate.)
In order to maximize the joint efficiency while minimizing dark
counts, a heralded detection scheme~\cite{Nam} was implemented
where one higher efficiency APD is triggered by one low
dark-count-rate SSPD. In this way the APD has far fewer
opportunities to generate dark counts, while still allowing us to
benefit from its higher quantum efficiency.

The NbN-meander SSPD that we employed in the experiment is a
nanoscale superconducting wire operated at a temperature of
$\simeq 3$\,K and biased close to its critical current. When a
photon strikes the wire, it forms a hot spot which momentarily
breaks the superconductivity and causes a transient voltage on the
device that is registered as a photon detection.  The entire
process happens very quickly, leading to a dead time of $\simeq
10$\,ns.  In order to use this signal detector as a herald of the
idler photon, we need to delay the idler photon in a ``delay
fiber'' to allow an electronic heralding trigger pulse to arrive
at the APD.  At the same time an electrical pulse is delayed
(using a Stanford Research Systems DG535) and sent to a photon
coincidence counter.
The triggered APD output is reshaped by a
field-programmable-gate-array board (not shown in Fig.~\ref{Expr})
and sent back to the same photon counter to be recorded as a
coincidence count. We scan the delay time $\tau$ given by the
delay generator to locate the ``coincidence peak" (corresponding
to $\tau=\tau_0$), where the count value is significantly higher
than its neighboring peaks, which indicates that each member of
the photon pair from the same optical pulse has been captured by
the SSPD/APD combination. This coincidence peak value is hereafter
referred to as (total) coincidences. Accidental coincidences,
mainly due to Raman noise and dark counts, are conveniently
recorded by setting $\tau=\tau_0-20$\,ns, where 20\,ns is the
period between consecutive pump pulses.

\begin{figure}[t]
\centering
\includegraphics[scale=.9]{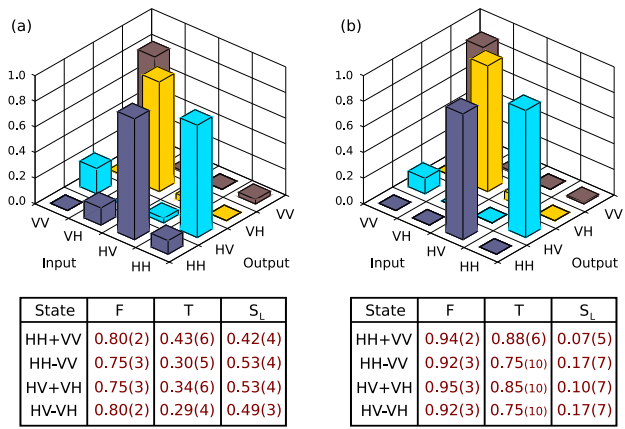} \vspace{-6pt}
\caption{(color online) (a) Experimentally measured truth table
for the \textsc{cnot} gate in the logical basis (data represent
coincidence counts minus accidental counts due to Raman noise and
detector dark counts). The highest peak values are: $0.81 \pm
0.05$ (${VV, VV}$), $0.85 \pm 0.06$ (${VH, VH}$), $0.94 \pm 0.07$
(${HV, HH}$), and $0.88 \pm 0.06$ (${HH, HV}$). Also shown is the
Bell-state characterization for the same gate. Fidelity ($F$),
tangle ($T \equiv$ Concurrence$^2$), and linear entropy ($S_L
\equiv \frac{4}{3}\left\{1-\textrm{Tr}(\rho^2)\right\}$) are given
for all \textsc{cnot}-generated Bell states. (b) The same truth
table and Bell-state data is shown after subtracting accidental
coincidences due to multiple photon-pair production. Peak values
are: $0.88 \pm 0.05$, $0.98 \pm 0.06$, $0.94 \pm 0.05$, and $0.99
\pm 0.05$.
}\label{results}
\vspace{-12pt}
\end{figure}

To characterize the performance of our \textsc{cnot} gate, we
first input four possible logical basis states for the control
($C$) and target ($T$) qubits: $|V\rangle_C |V\rangle_T$,
$|V\rangle_C |H\rangle_T$, $|H\rangle_C |V\rangle_T$, $|H\rangle_C
|H\rangle_T$, and record coincidences and accidental coincidences
for each output state in the above logical basis. We then subtract
the latter coincidences from the former to get true coincidences
for each case, which, after being transformed into probabilities,
are plotted in Fig.~\ref{results} (a), displaying the truth table
for the gate.
It can be seen from Fig.~\ref{results} (a) that the gate works
quite well in the logical basis, with an average fidelity
(probability of getting the correct output averaged over all
logical inputs) of 0.87 (87\%).
During the experiment, we apply a triangular-wave voltage to a
piezoelectric transducer (not shown in Fig.~\ref{Expr}) placed
within the translation stage under M1; this dithering
technique~\cite{Lu} effectively averages out single-photon
interference originating from the QS source due to the
experiment's high pump power and stabilizes the single-count
measurements recorded by the SSPD. Such a technique is not
necessary in the low pump-power regime, as single stray photons
become negligible.


As a demonstration of the entangling capability of the gate, we
send in four separable states $|D\rangle_C |V\rangle_T$,
$|A\rangle_C |V\rangle_T$, $|D\rangle_C |H\rangle_T$, and
$|A\rangle_C |H\rangle_T$ ($|D\rangle \equiv |H\rangle +
|V\rangle$ and $|A\rangle \equiv |H\rangle -|V\rangle$), which
theoretically should be transformed by the gate into four
maximally entangled Bell states: $|\Phi^{\pm}\rangle$,
$|\Psi^{\pm}\rangle$, respectively. We then characterize
each output state using quantum-state tomography, and reconstruct
its density matrix using the maximum likelihood
method~\cite{James,Altepeter}. Corrections are made on the final
results to account for small pump-power fluctuations during the
measurements. The reconstructed density matrix $\hat{\rho}$ for
each output state is then used to calculate its fidelity with its
theoretically predicted Bell state (e.g., $F_{\Psi^{-}} \equiv
\langle\Psi^{-}|\hat{\rho}|\Psi^{-}\rangle$), its tangle
\cite{Nielsen}, and its linear entropy
\cite{James,Munro,Nielsen}. The results are summarized in
Fig.~\ref{results} (a).

While mode mismatch explains some of the deviation from ideal
\textsc{cnot} performance shown in Fig.~\ref{results} (a), the
mode matching quality implied by the source's HOM
visibility~\cite{ChenPRL} only accounts for a degradation of
entangled fidelities to $\simeq 95\%$.  Imperfect optics account
for $\simeq 1\%$ of additional error. The major source of error
is, in fact, due to multiple-pair creations in the
identical-photon source. In comparison with Ref.~\cite{ChenPRL},
here we pump the system with relatively high pump power (total
average power $P=450\,\mu$W) to combat the gate's inherent 8/9
loss.
This pump power leads to $\simeq 15\%$ of all gate inputs arising
from multiple-pair events. In order to accurately characterize the
performance of the \textsc{cnot} gate on \emph{single} pairs of
input photons, we estimate the multiple-pair contributions to the
measurements, subtract them from the data, and show the resulting
\textsc{cnot} gate performance in Fig.~\ref{results} (b).  These
much higher fidelities and gate performance ($\simeq 95\%$
canonical-basis fidelity and $\simeq 93\%$ entangled fidelities)
are consistent with our previous error estimates.

To estimate the multiple-pair contributions to the data, we
perform a new type of maximum-likelihood process
tomography~\cite{OBrienPRL}. By assuming that only linear optics
are present within our gate, we model a process as the
transformations of each of four single-photon input modes (two
polarizations and two spatial modes) into six single-photon output
modes (the input modes plus the two PPBS dump ports). These four
transformations can then be used to predict the gate's output for
any input. Note that in order to minimize the search parameters,
we only conduct the search over processes that do not cause
decoherence, and thus the search is \emph{not} a complete process
tomography. However, we are confident that the resulting ``pure''
process can be used to estimate accidental coincidences from
multiple pairs since it predicts with high precision all of our
measured data (the average difference of the predicted and
measured data was only 1.0 standard deviations, and the process
fidelity between the predicted process and the ideal \textsc{cnot}
gate was $\simeq 95\%$).



In conclusion, we have demonstrated a telecom-band quantum
\textsc{cnot} gate and characterized it using a heralding scheme
based on a superconducting single-photon detector.
The authors would like to acknowledge support
by the NSF under Grant No.\ EMT- 0523975.

\end{document}